\documentclass[12pt]{article}
\usepackage{latexsym}
\usepackage{epsfig}
\begin{document}
\newcommand{\beq}{\begin{equation}}
\newcommand{\eeq}{\end{equation}}
\newcommand{\bear}{\begin{eqnarray}}
\newcommand{\eear}{\end{eqnarray}}
\newcommand{\Ka}{K\"ahler }
\newcommand{\nn}{\nonumber}
\newcommand{\e}{\epsilon}
\newcommand{\te}{\tilde{\epsilon}}
\newcommand{\htheta}{\hat{\theta}}
\newcommand{\pd}{\partial}
\newcommand{\ghe}[3]{$\stackrel{\textstyle #1}
{\scriptstyle (#2,#3)}$}
\newcommand{\lre}{\multicolumn{3}{c}{$\stackrel{*}
{\longleftrightarrow}$}}
\newcommand\ft[2]{{\textstyle\frac{#1}{#2}}}
\newcommand{\dr}{\raise.3ex\hbox{$\stackrel{\leftarrow}{\delta}$}{}}
\newcommand{\dl}{\raise.3ex\hbox{$\stackrel{\rightarrow}{\delta}$}{}} 
\newcommand{\dirac}{/\!\!\!\partial}

\begin{titlepage}
\begin{flushright} KUL-TF-99/15 \\ hep-th/9905117
\end{flushright}
\vskip 2.5cm
\begin{center}
{\Large \bf Batalin--Vilkovisky gauge--fixing of a chiral two--form in six dimensions} \\
\vskip 1.5cm
{\bf Chris Van Den Broeck$^1$ and  Kor Van Hoof$\, ^2$} \\
\vskip 0.5cm
{\small Instituut voor Theoretische Fysica, \\
Katholieke Universiteit Leuven, B-3001 Leuven, Belgium }
\end{center}
\vskip 4cm
\begin{center}
{\bf Abstract}
\begin{quote}
We perform the gauge--fixing of the theory of a chiral two--form boson in six dimensions 
starting from the action given by Pasti, Sorokin and Tonin. We use the 
Batalin--Vilkovisky formalism, introducing antifields and writing down an 
extended action satisfying the classical master equation. Then we gauge--fix 
the three local symmetries of the extended action in two different ways. \end{quote}
\end{center}
\vfill
\hrule width 5cm
\vskip 2mm
{\small $^1$ E-mail: chris.vandenbroeck@fys.kuleuven.ac.be \\
$^2$ E-mail: kor.vanhoof@fys.kuleuven.ac.be}
\end{titlepage}

\section{Introduction}

Chiral bosons are antisymmetric tensors of rank $2n$ in $4n+2$ dimensions with (anti)self--dual field strengths. Their description in terms of a manifestly Lorentz-covariant action has been a longstanding problem. The core of this problem is the first order self--duality condition. The absence of a Lorentz covariant action makes an analysis of quantum properties of these theories more cumbersome. In 6 dimensions, the chiral boson is a (anti)self--dual two--tensor. It appears in the worldvolume action of the $M5$--brane \cite{M5brane} or in 6--dimensional chiral  supergravity \cite{6dsugra}.
\par
Some proposals for Lorentz invariant Lagrangians involved squaring the second class self--duality constraint \cite{siegel} or introducing an infinite tower of auxiliary fields \cite{MCWY}. Their quantum properties were studied in \cite{MCWY, IS}. The problem with the infinite tower of auxiliary fields of \cite{MCWY} is to find a consistent truncation. This approach was generalised to higher order forms in \cite{DH}.
\par
By giving up explicit Lorentz invariance, it became possible to construct a new class of actions \cite{HT, Schwarz}. In \cite{Schwarz}, 6--dimensional actions were constructed with manifest 5-dimensional Lorentz symmetry. The proof of 6--dimensional Lorentz invariance needed a non-trivial check.
\par
By introducing {\sl one} auxiliary scalar field, Lorentz covariant actions for chiral $p$--forms were constructed in \cite{PST1, pst2, pst3, dlt}. This auxiliary scalar field appears non-polynomially in the action. The action is invariant under two new gauge symmetries that depend on this scalar. One of the gauge symmetries makes it possible to gauge away this new auxiliary field.  The relation to the approach with an infinite number of Lagrange multipliers \cite{DH} and the formulation with manifest 5--dimensional Lorentz invariance \cite{Schwarz} is explained in \cite{PST1}. Later, also $\kappa$-symmetric, Lorentz invariant world--volume actions were constructed for the 5--brane of $M$--theory \cite{pst2, pst3, CKVP}. 
\par
In this paper, we study the free self--dual two--tensor in 6 dimensions in the Batalin--Vilkovisky (BV) formalism. This free action is superconformally invariant and it is the gauge--fixed action of the $\kappa$--symmetric $M5$--brane worldvolume action, restricted to quadratic terms, as proven in \cite{CKVP}. Up to now, a description for interacting tensor multiplets is lacking, although some attempts have been made \cite{intten}.
\par 
The BV--formalism \cite{bvrev} is suited to study systems with different types of gauge symmetry structures, e.g. reducible gauge algebras. The classical BV--formalism leads to a gauge--fixed action. The quantum BV--formalism enables the calculation of anomalies. Here, we restrict ourselves to the classical analysis of the action of the chiral 2--form in 6 dimensions.
\par
We start from the Pasti-Sorokin-Tonin (PST) action \cite{PST1}. We build an extended action for the chiral tensor that satisfies the classical master equation. For this, we have to introduce ghosts for the gauge symmetries and ghosts for ghosts for the reducible symmetries. 
\par
Using different canonical transformations, we find gauge--fixed actions corresponding to two gauge choices. The first action is a new, fully covariant  one, while another gauge choice gives rise to the gauge--fixing of \cite{Schwarz}.
\par
The paper is organised as follows. In section~\ref{s:bvform}, a short review of the classical part of the BV--formalism is given. In section~\ref{s:extS}, the extended action is built for the self--dual 2--tensor using the BV--formalism. Two possible gauge--fixings of this extended action are given in section~\ref{s:gaugefix}. 

\section{Review of the Batalin--Vilkovisky formalism}\label{s:bvform}

In this section we give a brief account of the classical BV--procedure. More elaborate reviews can be found in \cite{bvrev}. We will emphasize the case where the gauge symmetries  of the classical action have zero modes, so that a second generation of ghosts will have to be introduced. A 
{\sl ghost number} $g$ is assigned to the $g$th generation of ghosts, where the 
classical fields have $g=0$. For every field, including the ghosts, an 
antifield with opposite statistics is introduced. We denote the physical and ghost fields by $\Phi^A$ and their antifield counterparts by $\Phi^\ast_A$. 
The antifields are assigned a ghost number such that
\beq
g(\Phi^A) + g(\Phi^\ast_A) = -1\, .
\eeq
We also define an {\sl antifield number}, which is zero for fields and equal to $-g$ for antifields. 
For functionals $F(\Phi,\Phi^\ast)$, $G(\Phi,\Phi^\ast)$, we 
introduce the {\sl antibracket}
\beq
(F,G) = F{{\dr}\over{\delta \Phi^A}} {{\dl}\over{\delta \Phi^\ast_A}}G
      - F{{\dr}\over{\delta \Phi^\ast_A}} {{\dl}\over{\delta \Phi^A}}G,
\label{antibracket}
\eeq
where summation over $A$ is understood. Sometimes we will add to the
antibracket an index between brackets: $(F,G)_{(n)}$ means that in (\ref{antibracket}), only the terms with antifield number $n$ are to 
be included. 

The {\sl minimal extended action} is defined by adding to the classical
action a part containing the antifields, multiplied by the BRST-transformations. Becchi, Rouet, Stora and Tuytin replaced the parameters of the gauge transformations into ghost fields:
\bear
S_{min} & = & S_{cl} + S^1 \nonumber\\
        & = & S_{cl} + \Phi^\ast_A R^A{}_B \, c^B,\label{S1}
\eear
where $\delta \Phi^A = R^A{}_B(\phi^C) \, \varepsilon^B$ is the infinitesimal gauge 
transformation with parameter $\varepsilon^B$. The extra term has antifield number $1$.

If the gauge transformations of $S_{cl}$ have zero modes, extra terms 
have to be introduced which have an analogous form: an antifield which
will be the antighost of the relevant gauge transformation, multiplied by
the zero mode transformation, where the parameter has been replaced by
a ghost for ghost. This yields a term of antifield number $2$. An 
additional term of the same antifield number is introduced for the 
commutators of the gauge symmetries, of the form
\beq\label{S2comm}
\int dx \ft 12 (-)^B c^\ast_A \, T^A_{BC} \, c^B c^C,
\eeq
where the $T^A_{BC}$ are the structure functions
associated to the commutators of the gauge symmetries and $(-)^B$ is the fermion number of ghost $c^B$.
Other terms may have to be added to the action until it satisfies the {\sl properness condition}.
This essentially states that for every gauge symmetry a ghost has been
introduced, and for every zero mode a ghost for ghost. This can be
checked by counting the number of zero modes of the Hessian
\beq
S_A{}^B={{\dl}\over{\delta \Phi^A}} S {{\dr}\over{\delta \Phi^\ast_B}}
\eeq
at the stationary surface. This is the surface where the classical fields
satisfy the field equations and the ghosts and antifields are set to zero. 
The result should be half the number of fields and antifields.

More terms of higher antifield number, $S^3,\ldots,S^n$, are added until 
the resulting action, called the {\sl extended action}, satisfies the 
{\sl classical master equation}, which is 
\beq
(S_{ext},S_{ext}) = 0.
\eeq
The truncation of the extended action to the antifield--independent part
should be the classical action. 
\par
{\sl Canonical transformations} from the basis 
$\{\Phi^A,\Phi^\ast_A\}$ to another set 
$\{\tilde{\Phi}^A,\tilde{\Phi}^\ast_A\}$ are defined to leave the antibrackets
invariant. Often they can be determined by a fermionic
generating function of ghost number $-1$, $F(\Phi,\tilde{\Phi}^\ast)$, 
such that
\bear
\tilde{\Phi}^A = {{\delta F(\Phi,\tilde{\Phi}^\ast)}
                     \over{\delta \tilde{\Phi}^\ast_A}}\, , &&
\Phi^\ast_A = {{\delta F(\Phi,\tilde{\Phi}^\ast)}
                     \over{\delta\Phi^A}}\, .
\eear
The generating function is of the form
\beq
F(\Phi,\tilde{\Phi}^\ast) = \Phi^A \tilde{\Phi}^\ast_A + \Psi(\Phi).
\eeq
The function $\Psi$, also a fermion of ghost number $-1$, is called the
{\sl gauge fermion}.

Gauge--fixing is performed by canonical transformations on the set of 
fields and antifields, such that the antifield--independent part of the
extended action has a Hessian that is invertible on the stationary surface.
In that case, well--defined propagators can be calculated.
To do the gauge--fixing, it will be necessary to introduce auxiliary fields and their antifields $b$ and
$b^\ast$. They should not have any influence on the master equation, and 
the physical content of the action should not be changed. For example,
one gives the field $b$ a ghost number $-1$, so that its antifield has 
ghost number zero, and one adds to the action the term $S_{nm} = (b^\ast)^2$.
Fields with negative ghost number are called {\sl non--minimal}.

For each gauge invariance one introduces such a non--minimal field $b^i$,
and gauge--fixing functions $F_i(\Phi_{cl})$, where the $\Phi^A_{cl}$ are
the classical fields. The gauge fermion is then taken to be
\beq
\Psi = b^i F_i(\Phi_{cl}).  
\eeq
One also has to take care of the zero modes of the gauge transformations.
Therefore, one has to introduce new auxiliary fields, but since they will
have to be fermionic, it will be impossible to add a term $(b^\ast)^2$
to the extended action. We will deal with this problem as it arises.

This concludes our summary of the BV--formalism. For justification of the
statements made here, and for further details, one should consult one 
of the references \cite{bvrev}.

\section{The extended action for a chiral $2$--form}\label{s:extS}

In this section, we first give the notations of the six--dimensional PST--action, and then build the extended action for the chiral 2--form introducing ghosts
and ghosts for ghosts associated to the gauge symmetries and their zero modes. Since we are interested in studying the gauge symmetries of the chiral tensor, we will restrict the action to the terms with the chiral two--tensor and not consider the full supersymmetric model.

\subsection{The classical action and its symmetries} 

The classical action for the self--dual tensor multiplet in $6$ flat dimensions in the PST--formalism \cite{PST1} is
\beq\label{Scl}
S_{cl} = \int d^6 x  -\ft 12 H^-_{ab} H^{\ast ab} 
\, .
\eeq
The action contains a tensor $B_{ab}$ and an auxiliary field $a$. This is the action for one self--dual tensor. The description for interacting tensors, the non--abelian generalisation of the free model (\ref{Scl}), is not known. To realise a self--dual tensor in a supersymmetric context, chiral supersymmetry in 6 dimensions is needed. The action of the self--dual tensor multiplet with rigid superconformal symmetry is presented in \cite{CKVP}. We will confine ourselves to the study of the gauge symmetries of the bosonic model (\ref{Scl}). The following notations are used:

\begin{eqnarray}
&& u_a=\partial_a a\, ,\qquad u^2=u^a u_a\, ,\nonumber \\
&& H=dB\, ,\qquad H_{abc}=3\partial_{[a}
B_{bc ]}\, ,\nonumber \\
&& H_{ab}={{u^c}\over{\sqrt{u^2}}} H_{abc}\, ,\nonumber \\
&& H^*_{ab}={{u^c}\over{\sqrt{u^2}}} H^*_{abc}\, ,\qquad \mbox{where } H^*_{abc}=\frac{1}{6}\epsilon_{abcdef} H^{def}\, , \nonumber \\
&& H^\pm_{ab}={u^c\over{\sqrt{u^2}}} H^\pm_{abc}\, , \qquad \mbox{where } H^\pm_{abc} = \ft 12 (H_{abc} \pm H^*_{abc})\, .
\label{defn}
\end{eqnarray}
This action has the following three gauge symmetries:
\bear 
\delta_I B_{ab} = 2 \pd_{[a} \Lambda_{b]}\, , && \delta_I a = 0\, , \nn \\
\delta_{II} B_{ab} = 2 H^-_{ab} {{\phi}\over{\sqrt{u^2}}}\, , 
&& \delta_{II} a = \phi\, , \nn \\
\delta_{III} B_{ab} = u_{[a} \psi_{b]}\, , && \delta_{III} a = 0\, .\label{sym} 
\eear
The first symmetry is the usual gauge symmetry for a $p$--form. The second and the third symmetry are `new' symmetries, enabled by the introduction of the scalar field $a$.
The first and the third symmetries are reducible. They have 3 zero modes:
\bear
(a) && \Lambda_a= \pd_a \Lambda\, , \nonumber \\
(b) && \psi_a = u_a \psi\, , \nonumber \\
(c) && \Lambda_a = u_a \Lambda'\, , \,\,\, \psi_a = 2 \pd_a \Lambda' \,.
\label{third}
\eear

\subsection{The extended action of the chiral 2--form}

Using the existence of the gauge symmetries and their reducibility conditions, we will build the extended action of the chiral 2--form. The first step to achieve an extended action is to introduce antifields $B^\ast_{ab}$ and $a^\ast$
for the classical fields, and ghosts $c_a$, $c$ and $c'_{b}$ associated
respectively to symmetries I, II and III. The ghost number and statistics of the different fields and antifields can be found in table~\ref{ghnstat}. \\
\begin{table}[ht]
\begin{center}
\begin{tabular}{|c|c|c|c|c|c|c|c|c|c|c|c|c|c|c|c|c|}\hline
$\Phi$&$B_{ab}$&$a$&$B_{ab}^*$&$a^*$&$c_a$&$c$&$c'_a$&$c_a^*$&$c^*$&$c_a^{'*}$&$d_1
$&$d_2$&$d_3$&$d_1^*$&$d_2^*$&$d_3^*$\\ \hline
$g(\Phi)$&0&0&-1&-1&1&1&1&-2&-2&-2&2&2&2&-3&-3&-3\\ \hline
\mbox{stat($\Phi$)}&+&+&-&-&-&-&-&+&+&+&+&+&+&-&-&-\\ \hline
\end{tabular}
\caption{Ghost number and statistics of the minimal fields and their antifields.
\label{ghnstat}}
\end{center}
\end{table}
The gauge symmetries yield the following contribution to 
the extended action at {\sl antifield number} 1 as given by (\ref{S1}):
\beq\label{S1gauge}
S^1 = \int d^6 x \left( B^{\ast ab} (2 \pd_a c_b + 2 H^-_{ab} 
{{c}\over{\sqrt{u^2}}} + u_a c'_b) + a^\ast c \right).
\eeq
A contribution at antifield number 2 comes from the ghosts for ghosts $d_1$, $d_2$ and $d_3$, 
associated to the 3 zero modes of the gauge symmetries (\ref{third}).
This gives
\beq\label{S2red}
S^2_1 = \int d^6 x \left( c^\ast_a (\pd^a d_1 + u^a d_3) 
                         +c'^\ast_a (u^a d_2  + 2\pd_a) d_3 \right),
\eeq
where $c^\ast_a$ and $c'^\ast_a$ are the antifields associated to the
ghosts $c_a$ and $c'_a$. In $S^2$, we also have to include a term related to the commutators of the 
symmetries as indicated in (\ref{S2comm}). These commutators are
\bear
\left[ \delta_{II}(\phi_2), \delta_{II}(\phi_1) \right]  
& = & \delta_{III} (4 \frac{H^-_{ab}}{(u^2)^{3/2}}  
(\phi_1 \pd^b \phi_2 - \phi_2 \pd^b \phi_1))\, , \\
\left[ \delta_{II}(\phi), \delta_{III}(\psi_a) \right] & = & \delta_I 
(\ft 12 \psi_a \phi)
+ \delta_{III}(2\pd_{[a} \psi_{c]}\cdot u^c {{\phi}\over{u^2}})\, . 
\eear
We get the additional term in the action
\beq\label{S2commut}
S^2_2 = \int d^6 x \left( \ft 12 c^\ast_a c'^a c - c'^\ast _a 
\pd^{[a} c'^{b]}\cdot {{u_b}\over{u^2}} c 
+ 4 c'^\ast_a {{H^{-ab}}\over{(u^2)^{3/2}}} \, c \pd_b c \right).
\eeq
It is straightforward to check that the action 
\beq
S = S_{cl} + S^1 + S^2, \label{actupto2}
\eeq
with $S^2=S^2_1+S^2_2$, satisfies the properness condition, i.e. 
that the Hessian
\beq
S_A{}^B = {{\dl}\over{\delta \Phi^A}} S {{\dr}\over{\delta \Phi^\ast_B}}
\eeq
at the stationary surface has a number of zero modes that is exactly
half its dimension. To check it, it is convenient to choose a point
on the surface where $H^-_{abc} = 0$. The number of zero modes of the 
Hessian cannot vary along the stationary surface.
\par
As far as antibrackets are concerned, one has 
\beq
(S_{cl},S^1) = 0\, .
\eeq
This is simply a consequence of the gauge invariance of the classical
action. However, the classical master equation is not yet satisfied; in
particular, one has the antibracket
\bear
2(S^1,S^2)_{(2)} + (S^2,S^2)_{(2)} & = & 8 c'^\ast_a {{H^-_{bcd}}\over{(u^2)^3}}
 u^a u^d \pd^c c \cdot c \pd^b c \nonumber\\
&& + c'^\ast_a \left(-2 {{u^a u^b}\over{u^2}} \pd_b d_2 \cdot c + 2 \pd^a (d_2 c)\right) \nonumber\\
&& + c^\ast_a \left(2 \pd^a (d_3 c) - u^a d_2 c \right )
+ 2 c'^\ast_a {{u^a u^b}\over{(u^2)^2}} \pd_{[c} c'_{b]} \cdot c \pd^c c.
\eear
Since this is not $0$, we have to add terms at antifield number 3. The choice that works is
\bear\label{S3}
S^3 &=& -d^\ast_1 d_3 c \nonumber \\
    && + d^\ast_2 \left(-4 {{H^-_{ab}}\over{(u^2)^{5/2}}} \pd^b c\cdot c\pd^a c
       +2 {{u^a}\over{(u^2)^2}} \pd_{[a} c'_{b]} \cdot c \pd^b c
       + {{u^a}\over{u^2}} \pd_a d_2 \cdot c \right) \nonumber \\
    && + d^\ast_3 d_2 c\, . 
\eear
Then 
\begin{eqnarray}
2(S^2,S^3)_{(2)} -2(S^1,S^2)_{(2)}+(S^2,S^2)_{(2)} & = & 0\, ,\nonumber \\
2(S^1,S^3)_{(3)} + 2(S^2,S^3)_{(3)} + (S^3,S^3)_{(3)} & = & 0\, .
\end{eqnarray}
We conclude that the sum of (\ref{Scl}), (\ref{S1gauge}), (\ref{S2red}), (\ref{S2commut}) and (\ref{S3}) is a good extended action. This action satisfies the classical master equation, the properness condition, and the classical limit (deleting all terms with non-zero ghost number in the action) gives the classical action (\ref{Scl}). So, all the conditions to have a good extended action are fulfilled.

\section{Gauge--fixings of the extended action}\label{s:gaugefix}

In this section, we present three different gauge--fixings. First, there is the
gauge--fixing that gives rise to the explicit self--duality condition of the 
2--form \cite{PST1}. Then a covariant gauge--fixing of the extended action is given. By allowing the gradient of the auxiliary scalar to point in an arbitrary direction, a non--covariant gauge--fixing, as used in
\cite{Schwarz, Lechner}, is found. 

\subsection{The self--duality of the two--form}

The field equation of the two--form $B_{ab}$ is
\begin{equation}\label{ea}
\epsilon^{lmnpqr}\partial_n({1\over{u^2}}u_p H^-_{qrs}u^s)=0\, .
\end{equation}
The most general solution of (\ref{ea}) is \cite{PST1}
\begin{equation}\label{gs}
H^-_{lmn}u^n = u^2\partial_{[l}\Phi_{m]}+u^n(\partial_n\Phi_{[l})u_{m]}+
u_{[l}(\partial_{m]}\Phi_n)u^n,
\end{equation}
This is a gauge transformation $III$ of $H_{lm}^-$. By using a Schouten identity, it can be proven that $H_{lm}^- =0 $ is equivalent to $H_{lmn}^- = 0$. This 
means that it is possible to find the self--duality of $B_{ab}$ by picking a gauge
choice for the third gauge transformation. So, the formulation with an auxiliary field gives rise to the explicit self--duality equation.

\subsection{A covariant gauge--fixing}

In this section we will present a covariant gauge--fixing of the extended action using gauge fermions in the BV--formalism. We start by gauge--fixing the three gauge symmetries (\ref{sym}) of the classical action. We will 
use the gauges
\bear
\pd_a B^{ab} &=& 0\, , \label{gaugeI} \\
u^2 &=& 1\, , \label{gaugeII}\\
u_a B^{ab} &=& 0\, . \label{gaugeIII}
\eear
(\ref{gaugeII}) is a Lorentz invariant gauge--fixing for symmetry II.
The gauge (\ref{gaugeIII}) fixes symmetry III and is the analogue of the 
Lorentz gauge; using a transformation III, one can remove the component of 
$B_{ab}$ that is parallel to the vector $u^a$.
\par
To facilitate the gauge--fixing of symmetry I, we first rewrite the 
classical action as
\beq 
S_{cl} = \int d^6 x ( -\ft{1}{24} H^{abc} H_{abc}
                           + \ft 12 H^{-ab} H^-_{ab} )\, .
\eeq
For each of these gauge--fixings, one introduces a new non--minimal set of fermionic fields 
and their antifields and adds to the action a term quadratic in the antifields. In table~\ref{tbl:nmfields}, they are denoted by $b_a$. 
\tabcolsep 1pt
\begin{table}[ht]
\begin{center}\begin{tabular}{cccccccccccccccccc}
 &  &   &   &   &   &   &   & &\ghe{B_{ab}, a}0{-1}&   &   &   &   &   &   &
 & \\
 &  &   &   &   &   &   &   &$\swarrow$ &   &   &   &   &   &   &   &   &
\\
 &  &   &   &   &   &   & \ghe{b_a, b'^a, b}{-1}0  &   &   &   &\ghe{c^a, c'^a,
 c}1{-2}&   &
 &   &   &   &   \\
 &  &   &   &   &   & $\swarrow$   &   &   &   & $\swarrow$   &   &   &
 &  &   &   &        \\
 &  &   &   &   & \ghe{b'^{a_1} =\{l, n, q\}}0{-1}  &  \lre     & \ghe{b_{a_1} =\{k, m, p\}}{-2}1  &
 &   &   &\ghe{d_1, d_2, d_3}2{-3}   &   &   &   &
\end{tabular}
\caption{The fields for the gauge--fixing of theories with
a reducible gauge algebra, with (ghost number, ghost number of antifield)
and a schematic indication of non--degeneracy conditions of the gauge
fixing and the connected antifields in the non--minimal extended action.\label{tbl:nmfields}}
\end{center}
\end{table}     \tabcolsep 6pt         
\par
The gauge--fixing is done by introducing a gauge fermion for each symmetry:
\bear
\Psi_1 & = & b_a \pd_b B^{ab} \\
\Psi_2 & = & b'_a u_b B^{ab}  \\
\Psi_3 & = & b (u^2 -1) \, ,
\eear
and adding to the action the non--minimal terms
\beq
S_{nm_1} = -\ft 14 b^*_a b^{*a} -\ft 14 b'^*_a b'^{*a} + b^{*2} \, .
\eeq
The gauge symmetries of the classical action also had three zero modes. Their gauge--fixing is done in general by introducing two extra sets of bosonic fields and their antifields 
for each zero mode. In the table~\ref{tbl:nmfields} they are denoted by $b'^{a_1}$ and $b_{a_1}$.
The arrow between them is used to indicate that one adds 
a non--minimal term to the action
which is a product of their antifields.
\par
For the gauge--fixing of the three reducible gauge symmetries, one introduces the
bosons ${k,m,p}$ and ${l,n,q}$.
The following gauge fermions can be used:
\bear
\Psi_4 & = & k \pd_b c^b + l \pd_a b^a \\
\Psi_5 & = & m u_a c'^a + n u_a b'^a \\
\Psi_6 & = & p (u_a c^a -2 \pd_a c'^a )+ q (u_a b^a -2 \pd_a b'^a)\, .
\eear
To the action the following non--minimal terms are added:
\beq
S_{nm_2} = k^* l^* + \ft 12 m^* n^* + p^* q^* \, .
\eeq

The antifield--independent part of the action becomes
\bear
S &=& \int d^6 x \Big[ \ft 18 B_{ab} \Box B_{ab} -\ft 12 H^{-ab}H^-_{ab}  
-\ft 14 q^2 u^2 -2q \, u^a\partial_a \, l \nonumber \\
&&- l\Box l +(u^2 -1)^2 +q\Box q + n \, u^a\partial_a \, q -\ft 14 n^2 u^2 
-\ft 14 u_b B^{ab} \, u^c B_{ac} \nonumber \\
&& -b'^a\partial_a u_b\cdot c^b -b'^b \, u^a \partial_a \, c_b -\ft 12 b'^b u^2 c'_b 
- b^a \Box c_a \nonumber \\
&& +2\partial^a b^b \cdot H^-_{ab} {c\over{\sqrt{u^2}}} - \ft 12 c'_b \,  u^a\pd_a \, b^b
+\ft 52 u_a b^a \, \pd^b c'_b +\ft 12 b^a\pd_bu_a\cdot c'^b \nonumber \\ 
&& u_a c^a \, u_b b^b +4\pd_a c'^a\cdot \pd_b b'^b + (q b^a +2b u^a +n b'^a)\, \pd_a c
\nonumber \\
&&+p\, u^a\pd_a \, d_1 + k\Box d_1 + p \, u^2 \, d_3 + d_3 \, u^a \pd_a \, k + \ft 12 p\, u_a c'^a
c + \ft 12 k\, \pd^a(c'^a c) \nonumber \\
&&-2d_2 \, u^a\pd_a \, p +m\, u^2 \, d_2 + 4p\Box d_3 + 2m \, u^a\pd_a \, d_3 + 2 \pd_a p\cdot \pd^{[a}
c'^{b]}\cdot {u_b\over u^2}c \nonumber \\
&& -8\pd_a p \cdot {H^{-ab}\over {(u^2)^{3/2}}} \, c\pd_b c + (p c^a + mc'^a) \, \pd_a c
  \Big]\, .
\eear
This is a covariant gauge--fixed action for the self--dual 2--form in 6 dimensions. In principle this can be used to derive gravitational anomalies as in \cite{Lechner, Witten}, but the presence of the auxiliary scalar in the propagators makes an analysis of anomalies using this action very hard. The non-covariant gauge--fixing of the next section can be used for this.

\subsection{A non--covariant gauge--fixing}

The non--covariant gauge--fixing of the second symmetry in this section corresponds to $a=x^5$ in \cite{Schwarz}. In \cite{Lechner}, the second gauge symmetry was fixed by imposing $a = n_a x^a$ where $n_a$ is a unit vector such that $u^2 =1$. Using the Faddeev--Popov
approach, it was proven that this gauge--fixing gives rise to the propagators 
postulated and used in \cite{Witten} to calculate the gravitational anomalies 
of a chiral $2$--form in $6$ dimensions.
\par
The second gauge symmetry can be fixed by a canonical transformation that cannot be generated by a gauge fermion:
\bear
a & \rightarrow& -b^* +n_a x^a \\\nonumber
a^* & \rightarrow& b\, .
\eear
This means that 
\bear
u_c &\rightarrow& -\pd_c b^* + n_c\\\nonumber
u^2 &\rightarrow& 1 -2n^a\pd_a b^* + (\pd b^*)^2\, .
\eear
Using the gauge fermions 
\bear
\psi_1 & = & b_a\pd_b B^{ab} \\
\psi_3  & = & b'_a n_b B^{ab} \\
\psi_4 & = & k\pd_b c^b + l\pd _b b^b \\
\psi_5 & = & m n_a c'^a + n n_a b'^a \\
\psi_6 & = & p\,(n_a c^a + 2 \pd_a c'^a) + q\,(n_a b^a + 2\pd_a b'^a)\, ,
\eear
 for the other symmetries, and the corresponding non--minimal terms in the action
$$
S_{nm} = -\frac{1}{4} b^{*a}b^*_a +\frac{1}{2} b'^{*a}b'^*_a -k^*l^* 
+\frac{1}{2}m^* n^* -\frac{1}{2}p^* q^*\, ,
$$
the other symmetries of the action are gauge--fixed.
The antifield-independent part of the action is:
\bear
S &=& \int d^6 x \Big[ \ft 18 B_{ab} \Box B_{ab} 
-\ft 12 H^{-abc}n_c \, H^-_{abd}n^d -b^a \Box c_a - b'^a \, n^b\pd_b \,c_a \nonumber  \\
&& +2b^a \pd^b H^-_{abc}\cdot n^c c + \ft 32 n^a b_a\, \pd^b c'_b 
-\ft 12 b^a \, n^b\pd_b \, c'_a -\ft 12 b'^a c'_a + bc \nonumber \\
&& + k \Box d_1 + k \, n^a\pd_a \, d_3 + \ft 12 k\pd^a(c'_a c) + m d_2 
+2 m\, n^a\pd_a \, d_3 + \ft 12 q \, n^a\pd_a \, l + \ft 14 l \Box l \nonumber \\
&&+\ft 12 n_b B^{ab} \, n^c B_{ac} + \ft 12 n^2 -2 n \, n_a \pd^a \,q -2q \Box q -\ft 12 n^a c_a \, n^b b_b -2 \pd_a c'^a \cdot \pd_b b'^b\nonumber \\
&&+p \, n^a\pd_a \, d_1 + p d_3 + \ft 12 p \, n^a c'_a c -2d_2 \, n^a\pd_a \, p +4p\Box d_3 + 2\pd^a p \cdot \pd_{[a} c'_{b]}\cdot n^b c \nonumber \\
&&-8\pd^a p \cdot H^-_{abc} n^c \, c\pd^b c
\Big]\, .
\eear

\section{Conclusions}

In this article, we were able to find the full extended action of the chiral two--form in six dimensions using an auxiliary field $a$. For this, we had to introduce ghosts for the three gauge symmetries of the classical action and ghosts for ghosts for the three zero modes and add corresponding terms to the classical action.
We showed that the expansion in antifield number of the action satisfying the classical master equation terminates at antifield number 3. 
Since this action satisfies the properness condition and gives the classical action in the classical limit, it is a good extended action.
\par
We treated two different gauge--fixings, a non-covariant one as well as a 
covariant one. We note in passing that the non-covariant gauge fixing allows 
for an easy counting of the on--shell degrees of freedom: starting with 16 
(15 for the tensor and 1 for the auxiliary scalar), one ends up with 3 degrees 
of freedom after successive gauge--fixings of the three gauge symmetries. 

The two gauge--fixing schemes were implemented within the BV--formalism. 
The first one resulted in a fully Lorentz covariant gauge--fixed action.  
These gauge--fixed actions lead to different propagators that can be used in calculations of gravitational anomalies. From the form of the actions we derived, it is clear that our second gauge--fixing, also used in \cite{Lechner}, is easiest because of the absence of the auxiliary field $a$ in the propagators.
However, in the BV--formalism, a cohomology can be defined \cite{bvrev}, such that anomalies derived for different gauge--fixings (like the covariant one) are in the same cohomology class.

\medskip
\section*{Acknowledgements}

The authors would like to thank Piet Claus, Kurt Lechner, Walter Troost and Antoine Van Proeyen for very valuable comments and discussions. This work was supported by the European Commission TMR programme ERBFMRX-CT96-0045.


\begin{thebibliography}{99}
\bibitem{M5brane} C.G. Callan, J. A. Harvey, A. Strominger, {\sl World brane actions for string solitons}, Nucl. Phys. {\bf B367}, 60 (1991); \\
R. G{\" u}ven, {\sl Black p-brane solutions of D=11 supergravity theory}, Phys. Lett. {\bf B276} (1992) 49;\\
G.W. Gibbons, G. T. Horowitz, and  P.K. Townsend, {\sl Higher-dimensional resolution of dilatonic black hole singularities}, Class. Quantum Grav. {\bf 12} (1995) 297, hep-th/9410073;\\
K. Becker and M. Becker, {\sl Boundaries in $M$-theory} Nucl. Phys. {\bf B472}, 221 (1996), hep-th/9602071;\\
I. Bandos, K. Lechner, A. Nurmagambetov, P. Pasti, D. Sorokin, and M. Tonin, {\sl Covariant action for the super-five-brane of $M$-theory}, Phys. Rev. Lett. {\bf 78} (1997) 4332, hep-th/9701149.
\bibitem{6dsugra} P.S. Howe, G. Sierra and P.K. Townsend, {\sl Supersymmetry
in six dimensions}, Nucl. Phys. {\bf B221} (1983) 331;\\
L.J. Romans, {\sl Self--duality for interacting fields: covariant field equations for six dimensional chiral supergravities}, Nucl. Phys. {\bf B276} (1986) 71;\\
E. Bergshoeff, E. Sezgin and A. Van Proeyen, {\sl Superconformal tensor calculus and matter couplings in six dimensions}, Nucl. Phys. {\bf B264} (1986) 653;\\
F. Riccioni, {\sl Tensor multiplets in six dimensional (2,0) supergravity}, Phys. Lett. {\bf B422} (1998) 126, hep-th/9712176;\\
E. Bergshoeff, E. Sezgin, A. Van Proeyen {\sl $(2,0)$ tensor multiplets and conformal supergravity in $D=6$}, Class. Quant. Grav. {\bf 16}, (1999) 3193, hep-th/9904085.
\bibitem{siegel} W. Siegel, {\sl Manifest Lorentz invariance sometimes requires nonlinearity}, Nucl. Phys. {\bf B238} (1984) 307.
\bibitem{MCWY} B. McClain, Y.S. Wu and F. Yu, {\sl Covariant quantization of chiral bosons and $OSP(1,1|2)$ symmetry}, Nucl. Phys. {\bf B343} (1990) 689.
\bibitem{IS}
C. Imbimbo and A. Schwimmer, {\sl The Lagrangian formulation of chiral scalars}, Phys. Lett. {\bf B193}, 455 (1987);\\
C. M. Hull, {\sl Covariant quantization of chiral bosons and anomaly cancellation}, Phys. Lett. {\bf B206}, 234 (1988);\\
J. M. F. Labastida and M. Pernici, {\sl On the BRST quantization of chiral bosons}, Nucl. Phys. {\bf B297}, 557 (1988);\\
L. Mezincescu and R. I. Nepomechie, {\sl Critical dimensions for chiral bosons}, Phys. Rev. {\bf D37}, 3067 (1988).
\bibitem{DH} F. P. Devecchi and M. Henneaux, {\sl Covariant path integral for chiral p--forms}, Phys. Rev. {\bf D45}, 1606 (1996), hep-th/9603031.
\bibitem{HT} R. Floreanini and R. Jackiw, {\sl Self--dual fields as charge density solitons}, Phys. Rev. Lett. {\bf 59} (1987) 1873;\\
F. Bastianelli and P. van Nieuwenhuizen, {\sl Chiral bosons coupled to supergravity}, Phys. Lett. {\bf B217} (1989) 98;\\
F. Bastianelli and P. van Nieuwenhuizen, {\sl Gravitational anomalies from the action for self--dual antisymmetric tensor fields in $4k+2$ dimensions}, Phys. Rev. Lett. {\bf 63} (1989) 728;\\
M. Henneaux and C. Teitelboim, {\sl Dynamics of chiral (self--dual) $p$--forms}, Phys. Lett. {\bf B206}
(1989) 650.
\bibitem{Schwarz}M. Perry, J. Schwarz, {\sl Interacting chiral gauge-fields in six dimensions and Born-Infeld theory}, Nucl. Phys. B489 (1997) 47, hep-th/9611065;\\ 
J. Schwarz, {\sl Coupling a self--dual tensor to gravity in six dimensions}, Phys. Lett. {\bf B395} (1997) 191, hep-th/9701008;\\
M. Aganagic, J. Park, C. Popescu, J. Schwarz, {\sl World-volume action of the $M$ theory five-brane}, Nucl. Phys. {\bf B496} (1997) 191, hep-th/ 9701166.
\bibitem{PST1} P. Pasti, D. Sorokin and M. Tonin, {\sl On Lorentz invariant actions for chiral $p$--forms}, Phys. Rev. {\bf D55} (1997) 6292, hep-th/9611100.
\bibitem{pst2} P. Pasti, D. Sorokin and M. Tonin, {\sl Covariant action for a D=11 five-brane with the chiral field}, Phys. Lett. {\bf B398} (1997) 41, hep-th/9701037.
\bibitem{pst3} Igor Bandos, Kurt Lechner, Alexei Nurmagambetov, Paolo Pasti, Dmitri Sorokin, Mario Tonin, {\sl On the equivalence of different formulations of the $M$ Theory five--brane}, Phys. Lett. {\bf B408} (1997) 135, hep-th/9703127.
\bibitem{dlt} G. Dall'Agata, K. Lechner, M. Tonin, {\sl Covariant actions for N=1, D=6 theories with chiral bosons}, Nucl. Phys. {\bf B512} (1998) 179, hep-th/9710127.
\bibitem{CKVP} P. Claus, R. Kallosh, A. Van Proeyen, {\sl M 5-brane and superconformal (0,2) tensor multiplet in six dimensions}, Nucl.Phys. {\bf B518} (1998) 117, hep-th/9711161.
\bibitem{intten} A. Strominger, {\sl Open $p$-branes}, Phys. Lett. {\bf B383} (1996) 44, hep-th/9512059;\\
E. Witten, {\sl Five--branes and $M$-theory on an orbifold}, Nucl. Phys. {\bf B463} (1996) 383, hep-th/9512219;\\
N. Seiberg, {\sl Notes on theories with 16 supercharges}, Nucl. Phys. Proc. Suppl. {\bf 67} (1998) 158 ; hep-th/9705117.
\bibitem{bvrev} I.A. Batalin and G. A. Vilkovisky, {\sl Gauge algebra and 
quantization}, Phys. Lett. {\bf B102} (1981) 27; \\
I.A. Batalin and G. A. Vilkovisky, {\sl Quantization of gauge theories with  
linearly dependent generators}, Phys. Rev. {D28} (1983), 2567;\\
A.  Van Proeyen, in Proc.  of the Conference {\sl Strings \& Symmetries 1991}, 
Stony Brook, May 20--25, 1991, eds.  N. Berkovits et al., (World Sc.  Publ. Co., Singapore, 1992), 388;\\
W. Troost and A. Van Proeyen, in {\em Strings 93}, proceedings of the
Conference in Berkeley, CA,  24-29 May 1993, eds. M.B. Halpern, G. Rivlis and 
A. Sevrin, (World Sc. Publ. Co., Singapore), 158, hep-th/9307126;\\
W. Troost and A. Van Proeyen, in {\em Strings and Symmetries}, Lecture Notes in Physics, Vol. 447, Springer-Verlag, eds. G. Aktas, C. Saclioglu, M. Serdaroglu, 183 (1995), hep-th/9410162;\\
J. Gomis, J. Paris, S. Samuel, {\sl Antibracket, antifields and gauge-theory quantization}, Phys. Rept. {\bf 259} (1995) 1, hep-th/9412228;\\
S. Weinberg, {\sl The Quantum Theory of Fields}, vol. II Cambridge University Press, 1996;\\
M. Grisaru, A. Van Proeyen, D. Zanon, {\sl Quantization of the complex linear superfield}, Nucl. Phys. {\bf B502} (1997) 345, hep-th/9703081.
\bibitem{Lechner} K. Lechner, {\sl Self--dual tensors and gravitational anomalies in $4n+2$ dimensions}, Nucl. Phys. {\bf B537} (1999) 361, hep-th/9808025.
\bibitem{Witten} L. Alvarez-Gaum\'e, E. Witten, {\sl Gravitational anomalies}, Nucl. Phys. {\bf B234} (1984) 269.
\end{thebibliography}
\end{document}